\documentclass[conference,10pt]{IEEEtran}
\IEEEoverridecommandlockouts

\usepackage{epsfig,rotating,setspace,latexsym,amsmath,epsf,amssymb,amsfonts,bm,theorem,cite,algorithm,graphicx,epsf,authblk,epstopdf,color,algpseudocode,bbm}

\newtheorem{corollary}{Corollary}
\newtheorem{definition}{Definition}
\newtheorem{theorem}{Theorem}
\newtheorem{lemma}{Lemma}
\newenvironment{Proof}[1]{\medskip\par\noindent{\bf Proof:\,}\,#1}{{\mbox{\,$\blacksquare$}\par}}

\allowdisplaybreaks

\begin{document}

\title{Age-Minimal Online Policies for Energy Harvesting Sensors with Incremental Battery Recharges\thanks{This research was supported in part by the National Science Foundation under Grants ECCS-1549881, ECCS-1647198, ECCS-1650299, CCF 14-22111, and CNS 15-26608.}}


\author[1]{Ahmed Arafa}
\author[2]{Jing Yang}
\author[3]{Sennur Ulukus}
\author[1]{H. Vincent Poor}
\affil[1]{\normalsize Electrical Engineering Department, Princeton University}
\affil[2]{\normalsize School of Electrical Engineering and Computer Science, Pennsylvania State University}
\affil[3]{\normalsize Department of Electrical and Computer Engineering, University of Maryland}


\maketitle

\begin{abstract}
A sensor node that is sending measurement updates regarding some physical phenomenon to a destination is considered. The sensor relies on energy harvested from nature to transmit its updates, and is equipped with a finite $B$-sized battery to save its harvested energy. Energy recharges the battery {\it incrementally} in units, according to a Poisson process, and one update consumes one energy unit to reach the destination. The setting is {\it online,} where the energy arrival times are revealed causally after the energy is harvested. The goal is to update the destination in a timely manner, namely, such that the long term average {\it age of information} is minimized, subject to energy causality constraints. The age of information at a given time is defined as the time spent since the latest update has reached the destination. It is shown that the optimal update policy follows a {\it renewal} structure, where the inter-update times are independent, and the time durations between any two consecutive events of submitting an update and having $k$ units of energy remaining in the battery are independent and identically distributed for a given $k\leq B-1$. The optimal renewal policy for the case of $B=2$ energy units is explicitly characterized, and it is shown that it has an {\it energy-dependent threshold} structure, where the sensor updates only if the age grows above a certain threshold that is a function of the amount of energy in its battery.
\end{abstract}

\section{Introduction}

An energy harvesting sensor monitors some physical phenomenon and sends measurement updates about it to a destination. Updates are to be sent such that the long term average {\it age of information} is minimized. The age of information is the time spent since the freshest update has reached the destination. The sensor relies on energy harvested from nature to measure and send its updates, and is equipped with a finite $B$-sized battery to save its incoming energy. We characterize optimal {\it online} policies for this problem, where the sensor has only causal knowledge of the energy harvesting process.

In this work, we connect results from the energy harvesting communication literature and the age of information minimization literature by using the age of information metric as a means to assess the performance of a single-user energy harvesting communication channel. The energy harvesting communication literature is broadly categorized into offline and online settings, depending on whether the energy arrival times/amounts are known prior to the start of communication. Offline energy management works consider, e.g., single-user channels \cite{jingP2P, kayaEmax, omurFade, ruiZhangEH}; multiuser channels \cite{jingBC, omurBC, elifBC, jingMAC, kaya-interference}; and multi hop and relay channels \cite{ruiZhangRelay, gunduz2hop, berkDiamond-jour, varan_twc_jour, arafa_baknina_twc_dec_proc}. Recent online works include the near-optimal results for single-user and multiuser channels \cite{ozgur_online_su, ozgur_online_mac, baknina_online_mac, baknina_online_bc}, systems with processing costs \cite{baknina-online-proc}, and systems with general utilities \cite{arafa-baknina-gnrl-online}.

Age of information minimization is generally studied in a queuing-theoretic framework, including a single source setting \cite{yates_age_1}; multiple sources \cite{yates_age_mac}; variations of the single source setting such as randomly (out of order) arriving updates \cite{ephremides_age_random}, update management and control \cite{ephremides_age_management}, and nonlinear age metrics \cite{ephremides_age_non_linear, sun-age-mdp}; multi hop networks \cite{shroff_age_multi_hop}; broadcasting, multicasting, and multi streaming \cite{modiano-age-bc, soljanin-age-multicast, najm-age-multistream}; coding over erasures \cite{yates-age-erase-code}; and caching systems \cite{yates-age-cache}.

Assessing the performance of energy harvesting communication systems by the age of information metric has recently gained some attention \cite{yates_age_eh, elif_age_eh, arafa-age-2hop, arafa-age-var-serv, elif-age-Emax, jing-age-online, arafa-age-rbr, baknina-updt-info}. Except for \cite{arafa-age-var-serv}, an underlying assumption in these works is that energy expenditure is normalized, i.e., it takes one energy unit to send an update to the destination. References  \cite{yates_age_eh, elif_age_eh} study a system with an infinite-sized battery, with \cite{yates_age_eh} considering online scheduling with random service times (time for the update to take effect), and \cite{elif_age_eh} considering offline and online scheduling with zero service times. The offline policy in \cite{elif_age_eh} is extended to fixed non-zero service times in \cite{arafa-age-2hop} for single and multi hop settings, and to energy-controlled service times in \cite{arafa-age-var-serv}. The online policy in \cite{elif_age_eh} is found by dynamic programming in a discrete-time setting, and was shown to be of a threshold structure, where an update is sent only if the age of information is higher than a certain threshold. Motivated by the results in the infinite battery case, \cite{elif-age-Emax} then analyzes the performance of threshold policies under a finite-sized battery and varying channel assumptions, yet with no claim of optimality. Reference \cite{jing-age-online} proves the optimality of threshold policies when the battery size is equal to one unit using tools from renewal theory; it also provides an asymptotically optimal update policy when the battery size grows infinitely large. In our recent work \cite{arafa-age-rbr}, we extend the results of \cite{jing-age-online} and formally prove the optimality of threshold policies for any finite-sized battery in an online setting where the battery is randomly {\it fully} recharged over time, i.e., whenever energy is harvested, it completely fills up the battery. An interesting result is recently reported in \cite{baknina-updt-info}, where status updates send information, other than that related to measurements, in an energy harvesting single-user channel.

In this work, we complement our results in \cite{arafa-age-rbr} and study age-optimal online policies for an energy harvesting sensor with a finite battery with random {\it incremental} battery recharges; that is, energy is harvested in units as in \cite{elif-age-Emax, jing-age-online}, as opposed to full chunks as in \cite{arafa-age-rbr}. We extend the unit battery results of \cite{jing-age-online} and show that for a finite battery of size $B$, the optimal status update policy that minimizes the long term average age of information is a {\it renewal policy:} the times in between the two consecutive events where the sensor sends an update and has $k$ energy units remaining in its battery, for some $0\leq k\leq B-1$, are independent and identically distributed (i.i.d.). Further, we show that inter-update times are independent. Based on these results, we explicitly solve the case of $B=2$ energy units, and formally prove, using optimization tools, that the optimal policy is an {\it energy-dependent threshold policy:} the sensor submits an update only if the instantaneous age of information is above a certain threshold that depends on the energy in its battery.

\section{System Model and Problem Formulation}

We consider a sensor node that collects measurements from a physical phenomenon and sends updates to a destination over time. The sensor relies on energy harvested from nature to acquire and send its updates, and is equipped with a battery of finite size $B$ to save its incoming energy. The sensor consumes one unit of energy to measure and send out an update to the destination. We assume that updates are sent over an error-free link with negligible transmission times as in \cite{elif_age_eh, elif-age-Emax, jing-age-online, arafa-age-rbr}. Energy arrives (is harvested) one unit at a time, at times $\{t_1,t_2,\dots\}$ according to a Poisson process of rate $1$. Our setting is online in which energy arrival times are revealed causally over time; only the arrival rate is known a priori.

Let $s_i$ denote the time at which the sensor acquires (and transmits) the $i$th measurement update, and let $\mathcal{E}(t)$ denote the amount of energy remaining in the battery at time $t$. We then have the following energy causality constraint \cite{jingP2P}
\begin{align} \label{eq_en_caus}
\mathcal{E}\left(s_i^-\right)\geq1,\quad\forall i
\end{align}
We assume that we begin with an empty battery at time $0$, and that the battery evolves as follows over time
\begin{align} \label{eq_battery}
\mathcal{E}\left(s_i^-\right)=\min\left\{\mathcal{E}\left(s_{i-1}^-\right)-1+\mathcal{A}\left(x_i\right),B\right\}
\end{align}
where $x_i\triangleq s_i-s_{i-1}$, and $\mathcal{A}(x_i)$ denotes the amount of energy harvested in $[s_{i-1},s_i)$. Note that $\mathcal{A}(x_i)$ is a Poisson random variable with parameter $x_i$. We denote by $\mathcal{F}$, the set of feasible transmission times $\{s_i\}$ described by (\ref{eq_en_caus}) and (\ref{eq_battery}) in addition to an empty battery at time 0, i.e., $\mathcal{E}(0)=0$.

\begin{figure}[t]
\center
\includegraphics[scale=1]{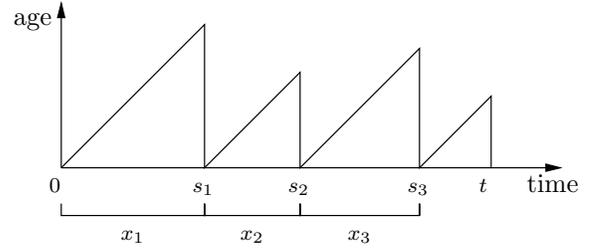}
\caption{Example of the age evolution versus time with $n(t)=3$.}
\label{fig_age_xmpl}
\end{figure}

The goal is to choose an online feasible transmission policy $\{s_i\}$ (or equivalently $\{x_i\}$) such that the long term average of the age of information experienced at the destination is minimized. The age of information is defined as the time elapsed since the latest update has reached the destination. The age at time $t$ is formally defined as
\begin{align}
a(t)\triangleq t-u(t)
\end{align}
where $u(t)$ is the time stamp of the latest update received before time $t$. Let $n(t)$ denote the total number of updates sent by time $t$. We are interested in minimizing the area under the age curve, see Fig.~\ref{fig_age_xmpl} for a possible sample path with $n(t)=3$. At time $t$, this area is given by
\begin{align} \label{eq_aoi}
r(t)\triangleq\frac{1}{2}\sum_{i=1}^{n(t)}x_i^2+\frac{1}{2}\left(t-s_{n(t)}\right)^2
\end{align}
and therefore the goal is to characterize the following quantity
\begin{align} \label{opt_main}
\bar{r}\triangleq\min_{{\bm x}\in\mathcal{F}}\limsup_{T\rightarrow\infty}\frac{1}{T}\mathbb{E}\left[r(T)\right]
\end{align}
where $\mathbb{E}(\cdot)$ is the expectation operator. In the next section, we characterize the structure of the optimal policy.

\section{Optimal Solution Structure:\\Renewal Type Policies}

In this section, we show that the optimal update policy that solves problem (\ref{opt_main}) has a renewal structure. Namely, we show that it is optimal to transmit updates in such a way that the inter-update delays are independent over time; and that the time durations in between the two consecutive events of transmitting an update and having $k\leq B-1$ units of energy left in the battery are i.i.d., i.e. these events occur at times that constitute a renewal process. We first introduce some notation.

Let the pair $\left(\mathcal{E}(t),a(t)\right)$ represent the state of the system at time $t$. Fix $k\in\{0,1,\dots,B-1\}$, and consider the state $(k,0)$, which means that the sensor has just submitted an update and has $k$ units of energy remaining in its battery. Let $l_i$ denote the time at which the system visits $(k,0)$ for the $i$th time. We use the term {\it epoch} to denote the time in between two consecutive visits to $(k,0)$. Observe that there can possibly be an infinite number of updates occurring in an epoch, depending on the energy arrival pattern and the update time decisions. For instance, in the $i$th epoch, which starts at $l_{i-1}$, one energy unit may arrive at some time $l_{i-1}+\tau_{1,i}$, at which the system goes to state $(k+1,\tau_{1,i})$, and then the sensor updates afterwards to get the system state back to $(k,0)$ again. Another possibility (if $k\geq1$) is that the sensor first updates at some time $l_{i-1}+x_{k,i}$, at which the system goes to state $(k-1,0)$, and then two consecutive energy units arrive at times $l_{i-1}+\tau_{1,i}$ and $l_{i-1}+\tau_{1,i}+\tau_{2,i}$, respectively, at which the system goes to state $(k+1,\tau_{1,i}+\tau_{2,i})$, and then the sensor updates afterwards to get the system state back to $(k,0)$ again. Depending on how many energy arrivals occur in the $i$th epoch, how far apart from each other they are, and the status update times, one can determine the length of the $i$th epoch and how many updates it has. Observe that the update policy in the $i$th epoch may depend on the history of events (energy arrivals and transmission updates) that occurred in previous epochs, which we denote by $\mathcal{H}_{i-1}$. Our main result in this section shows that this is not the case, under some mild technical conditions, and that epoch lengths should be i.i.d. We first have the following definition.

\begin{definition}[Uniformly Bounded Policy] \label{def_ubp}
An online policy whose inter-update times, as a function of the energy arrival times, have a bounded second moment.
\end{definition}

We focus on uniformly bounded policies as per Definition~\ref{def_ubp}. Such policies were also considered in \cite{jing-age-online} in the analysis of the $B=1$ case. We now have the following theorem; the proof is in Appendix~\ref{apndx_rnwl}.

\begin{theorem} \label{thm_rnwl}
In the optimal solution of problem (\ref{opt_main}), any uniformly bounded policy is a renewal policy. That is, the sequence $\{l_i\}$ denoting the times at which the system visits state $(k,0)$ forms a renewal process.
\end{theorem}

Based on Theorem~\ref{thm_rnwl}, the following corollary now follows.

\begin{corollary} \label{thm_indp}
In the optimal solution of problem (\ref{opt_main}), the inter-update times are independent.
\end{corollary}

\begin{Proof}
Observe that whenever an update occurs the system enters state $(j,0)$ for some $j\leq B-1$. The system then starts a new epoch {\it with respect to state} $(j,0)$. Since the choice of $k$ energy units in Theorem~\ref{thm_rnwl} is arbitrary, the results of the theorem now tell us that the update policy in that epoch, and therefore its length, is independent of the past history, in particular the past inter-update lengths.
\end{Proof}

In the next section, we show how to use the results of Theorem~\ref{thm_rnwl} and Corollary~\ref{thm_indp} to provide an explicit solution for the case of $B=2$ energy units.

\section{The Case $B=2$}

\begin{figure}[t]
\center
\includegraphics[scale=.9]{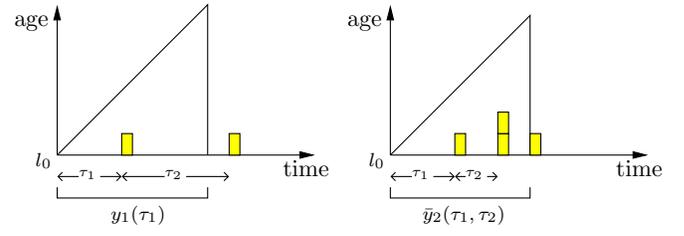}
\caption{Age of information versus time under the two possible ways of updating starting from state $(0,0)$ at time $l_0$. On the left, the second energy arrival occurs late, and hence we have one energy arrival followed by one update, returning to state $(0,0)$ again at $l_0+y_1(\tau_1)$. On the right, the second energy arrival occurs early, and hence we have two energy arrivals followed by one update, entering state $(1,0)$ at $l_0+\bar{y}_2(\tau_1,\tau_2)$. The yellow boxes represent energy units in the battery.}
\label{fig_age_xpln_1}
\end{figure}

Based on Corollary~\ref{thm_indp}, we now introduce the following notation regarding the update policy in a given epoch. Starting from state $(0,0)$ at time $l_0$, the sensor has to wait for the first energy arrival in the epoch, which occurs after some time $\tau_1$, and at which the system state becomes $(1,\tau_1)$. Since the sensor now has energy, it schedules its next update at $l_0+y_1(\tau_1)$, for some function $y_1(\cdot)$ to be optimally characterized. Now if another energy arrival occurs at time $l_0+\tau_1+\tau_2$, with $\tau_2>y_1(\tau_1)-\tau_1$, the sensor transmits the update as scheduled at $l_0+y_1(\tau_1)$ and the system state returns to $(0,0)$ again. On the other hand, if this second energy arrival occurs relatively early, i.e., $\tau_2\leq y_1(\tau_1)-\tau_1$, the system state becomes $(2,\tau_1+\tau_2)$ at $l_0+\tau_1+\tau_2$, and the sensor {\it reschedules} its update to be at $l_0+\bar{y}_2(\tau_1,\tau_2)$ instead of $l_0+y_1(\tau_1)$. Note that it is not clear so far whether $\bar{y}_2(\tau_1,\tau_2)$ depends only on the age $\tau_1+\tau_2$; we leave it as a general function of the pair $(\tau_1,\tau_2)$ for now. The above two cases are illustrated in Fig.~\ref{fig_age_xpln_1}.

\begin{figure}[t]
\center
\includegraphics[scale=.9]{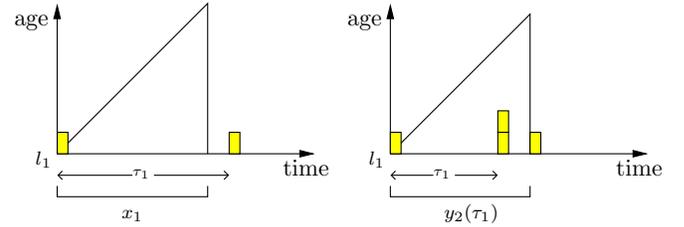}
\caption{Age of information versus time under the two possible ways of updating starting from state $(1,0)$ at time $l_1$. On the left, the first energy arrival occurs late, and hence the sensor updates and enters state $(0,0)$ at $l_1+x_1$. On the right, the first energy arrival occurs early, and hence we have one energy arrival followed by an update, returning to state $(1,0)$ at $l_1+y_2(\tau_1)$. The yellow boxes represent energy units in the battery.}
\label{fig_age_xpln_2}
\end{figure}

Once the sensor has two energy units in its battery, it will eventually send an update making the system state become $(1,0)$ at some time $l_1$. The sensor then schedules its next update at $l_1+x_1$, for some $x_1$ to be optimally characterized. If the first energy arrival after $l_1$ occurs at time $l_1+\tau_1$ with $\tau_1>x_1$, the sensor transmits the update at $l+x_1$ as scheduled, whence the state becomes $(0,0)$. Note that by energy causality, $x_1$ cannot depend on $\tau_1$, and since it also does not depend on the past history before $l_1$ (by Corollary~\ref{thm_indp}), it is therefore a constant. On the other hand, if the first energy arrival occurs relatively early, i.e., $\tau_1\leq x_1$, the state becomes $(2,\tau_1)$ at $l_1+\tau_1$, and the sensor {\it reschedules} the update to be at $l_1+y_{2}(\tau_1)$ instead of $l_1+x_1$. Note that it is not clear so far whether $y_2(\cdot)$ and $\bar{y}_2(\cdot,\cdot)$ are identical, since the former depends on only one random variable, as opposed to depending on two random variables in the latter; we optimally characterize both functions later on in the analysis. The above two cases are illustrated in Fig.~\ref{fig_age_xpln_2}.

In summary, the optimal update policy for the case $B=2$ in a given epoch is completely characterized by the constant $x_1$, and the functions $y_1(\cdot)$, $y_2(\cdot)$, and $\bar{y}_2(\cdot,\cdot)$. Since these represent the possible inter-update delays, we conclude by Corollary~\ref{thm_indp} that they do not depend on each other. We denote by $R\left(x_1,y_1,y_2,\bar{y}_2\right)$ and $L\left(x_1,y_1,y_2,\bar{y}_2\right)$ the area under the age curve in the epoch and its length, respectively, as a function of the policy $\left(x_1,y_1,y_2,\bar{y}_2\right)$. By Theorem~\ref{thm_rnwl} (and Corollary~\ref{thm_indp}), one can use the strong law of large numbers of renewal processes \cite{ross_stochastic} to reduce problem (\ref{opt_main}) to be an optimization over a single epoch as follows
\begin{align} \label{opt_rnwl}
\min_{x_1,y_1,y_2,\bar{y}_2} \quad &\frac{\mathbb{E}\left[R\left(x_1,y_1,y_2,\bar{y}_2\right)\right]}{\mathbb{E}\left[L\left(x_1,y_1,y_2,\bar{y}_2\right)\right]} \nonumber \\
\mbox{s.t.}~~~ \quad &x_1\geq0 \nonumber \\
&y_1(\tau)\geq\tau, \quad \forall \tau \nonumber \\
&y_2(\tau)\geq\tau, \quad \forall \tau \nonumber \\
&\bar{y}_2(\tau_1,\tau_2)\geq\tau_1+\tau_2, \quad \forall \tau_1,\tau_2
\end{align}
where the expectation is on the energy arrival patterns in the epoch. Note that the constraints on the functions $y_1$, $y_2$, and $\bar{y}_2$, represent the energy causality constraints. Next, in order to evaluate the expectations in the objective function, one needs to study the different patterns that can occur in a single epoch. We do so in the following subsection.

\subsection{Renewal State Analysis}

\begin{figure*}[t]
\center
\includegraphics[scale=1]{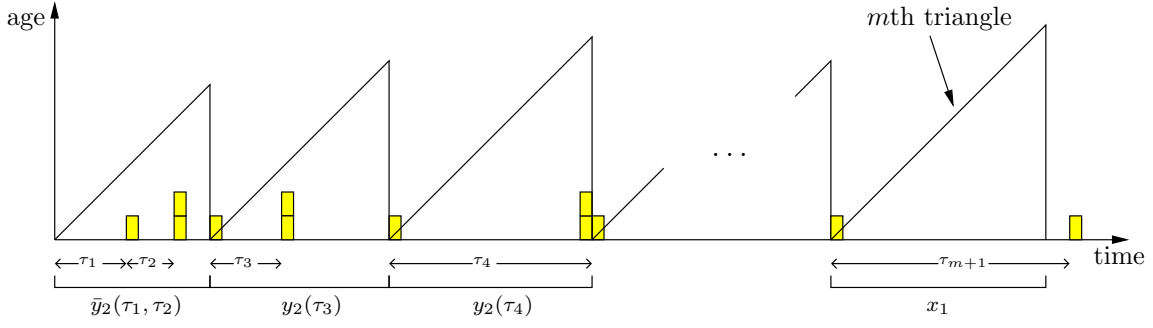}
\caption{General $m$th possible way to return to state $(0,0)$, $m\geq3$.}
\label{fig_age_way_m}
\end{figure*}

Consider the state $(0,0)$ as the renewal state\footnote{From Theorem~\ref{thm_rnwl}, we know that both states $(0,0)$ and $(1,0)$ are renewal states. While we choose to perform our analysis using state $(0,0)$, we note that one can reach the same results if state $(1,0)$ is chosen instead.}, and without loss of generality assume that we start at time $0$. Let us now state the possible ways of returning to that state. Note that the sensor has to wait for at least one energy arrival to update since it starts with no energy at state $(0,0)$.
\begin{itemize}
\item The first way to return to $(0,0)$ is to receive an energy arrival after $\tau_1$ time units, and then update at $y_1(\tau_1)$. This could happen if and only if the following energy arrival, occurring at $\tau_2$ time units after the first arrival, arrives after $y_1(\tau_1)-\tau_1$. See Fig.~\ref{fig_age_way_1}.

\begin{figure}[t]
\center
\includegraphics[scale=1]{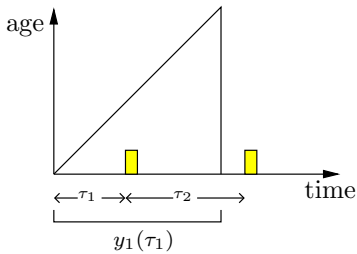}
\caption{First possible way to return to state $(0,0)$.}
\label{fig_age_way_1}
\end{figure}

\item The second way is to receive another energy arrival after the first one, before using the first energy unit to update. Then, submit the first update at $\bar{y}_2(\tau_1,\tau_2)$, which makes the state become $(1,0)$, and then submit another update after $x_1$ time units. This could happen if and only if the following energy arrival, occurring at $\tau_3$ time units after the first update, is such that $\tau_3>x_1$. See Fig.~\ref{fig_age_way_2}.

\begin{figure}[t]
\center
\includegraphics[scale=1]{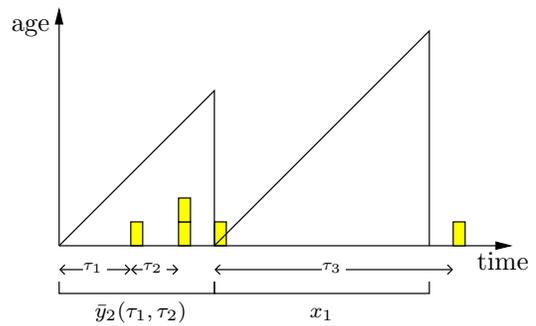}
\caption{Second possible way to return to state $(0,0)$.}
\label{fig_age_way_2}
\end{figure}

\item The third way is exactly as the second way, but with $\tau_3\leq x_1$, and hence the system goes to state $(2,\tau_3)$ with the third energy arrival. Then, the sensor updates after $y_2(\tau_3)$ time units from the first update (as opposed to $x_1$ in the second way), which makes the state become $(1,0)$, and then finally submit a third update after $x_1$ time units. As before, this could happen if and only if the following energy arrival, occurring at $\tau_4$ time units after the second update, is such that $\tau_4>x_1$. See Fig.~\ref{fig_age_way_3}.

\begin{figure}[t]
\center
\includegraphics[scale=.9]{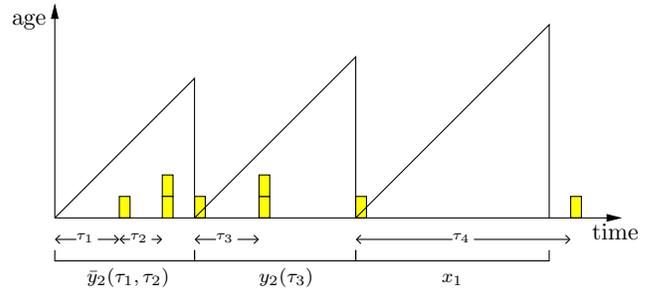}
\caption{Third possible way to return to state $(0,0)$.}
\label{fig_age_way_3}
\end{figure}

\item In general, the $m$th way, $m\geq3$, begins exactly as in the second way by submitting the first update at $\bar{y}_2(\tau_1,\tau_2)$. Then, the second phase of the third way, namely, going from state $(1,0)$ to $(2,\tau_3)$ to $(1,0)$ again, keeps repeating for $m-2$ times. By the end of these repetitions the system will be in state $(1,0)$. This is finally followed by the $m$th (and last) update after $x_1$ time units. See Fig.~\ref{fig_age_way_m}.
\end{itemize}

\begin{figure*}
\begin{align} 
R=&\frac{1}{2}y_1(\tau_1)^2\mathbbm{1}_{\tau_2>y_1(\tau_1)-\tau_1} 
+\left(\frac{1}{2}\bar{y}_2(\tau_1,\tau_2)^2+\frac{1}{2}x_1^2\right)\mathbbm{1}_{\tau_2\leq y_1(\tau_1)-\tau_1}\mathbbm{1}_{\tau_3>x_1} \nonumber \\
&+\left(\frac{1}{2}\bar{y}_2(\tau_1,\tau_2)^2+\frac{1}{2}y_2(\tau_3)^2+\frac{1}{2}x_1^2\right)\mathbbm{1}_{\tau_2\leq y_1(\tau_1)-\tau_1}\mathbbm{1}_{\tau_3\leq x_1}\mathbbm{1}_{\tau_4>x_1} \nonumber \\
&+\left(\frac{1}{2}\bar{y}_2(\tau_1,\tau_2)^2+\frac{1}{2}y_2(\tau_3)^2+\frac{1}{2}y_2(\tau_4)^2+\frac{1}{2}x_1^2\right)\mathbbm{1}_{\tau_2\leq y_1(\tau_1)-\tau_1}\mathbbm{1}_{\tau_3\leq x_1}\mathbbm{1}_{\tau_4\leq x_1}\mathbbm{1}_{\tau_5> x_1} \nonumber \\
&+\dots \label{eq_r_b2} \\
L=&y_1(\tau_1)\mathbbm{1}_{\tau_2>y_1(\tau_1)-\tau_1} 
+\left(\bar{y}_2(\tau_1,\tau_2)+x_1\right)\mathbbm{1}_{\tau_2\leq y_1(\tau_1)-\tau_1}\mathbbm{1}_{\tau_3>x_1} \nonumber \\
&+\left(\bar{y}_2(\tau_1,\tau_2)+y_2(\tau_3)+x_1\right)\mathbbm{1}_{\tau_2\leq y_1(\tau_1)-\tau_1}\mathbbm{1}_{\tau_3\leq x_1}\mathbbm{1}_{\tau_4>x_1} \nonumber \\
&+\left(\bar{y}_2(\tau_1,\tau_2)+y_2(\tau_3)+y_2(\tau_4)+x_1\right)\mathbbm{1}_{\tau_2\leq y_1(\tau_1)-\tau_1}\mathbbm{1}_{\tau_3\leq x_1}\mathbbm{1}_{\tau_4\leq x_1}\mathbbm{1}_{\tau_5> x_1} \nonumber \\
&+\dots \label{eq_l_b2}
\end{align}
\hrulefill
\end{figure*}

Based on the above, one can write the area under the age curve, $R$, in a single epoch as in equation (\ref{eq_r_b2}) at the top of the next page\footnote{From now onwards, we drop the dependency on the tuple $\left(x_1,y_1,y_2,\bar{y}_2\right)$ from $R$ and $L$ for convenience.}. There, $\mathbbm{1}_A=1$ if event $A$ is true, and is $0$ otherwise. Taking expectations and simplifying (mainly through using the fact that $\tau_i$'s are i.i.d.), we get
\begin{align} \label{eq_exp_r}
\mathbb{E}\left[R\right]=&\left(\frac{1}{2}x_1^2+e^{x_1}\int_0^{x_1}\frac{1}{2}y_2(\tau)^2e^{-\tau}d\tau\right) \nonumber\\
&\times\left(1-\int_0^\infty e^{-y_1(\tau)}d\tau\right)+\int_0^\infty \frac{1}{2}y_1(\tau)^2e^{-y_1(\tau)}d\tau \nonumber \\
&+\int_{\tau_1=0}^\infty\int_{\tau_2=0}^{y_1(\tau_1)-\tau_1}\frac{1}{2}\bar{y}_2(\tau_1,\tau_2)^2e^{-\tau_1}e^{-\tau_2}d\tau_1d\tau_2
\end{align}
Equation (\ref{eq_exp_r}) is justified in Appendix~\ref{apndx_jstfy}. Similarly, the epoch length $L$ is given by (\ref{eq_l_b2}) at the top of this page, and its expectation is given by
\begin{align} \label{eq_exp_l}
\mathbb{E}\left[L\right]=&\left(x_1+e^{x_1}\int_0^{x_1}y_2(\tau)e^{-\tau}d\tau\right) \nonumber \\
&\times\left(1-\int_0^\infty e^{-y_1(\tau)}d\tau\right)+\int_0^\infty y_1(\tau)e^{-y_1(\tau)}d\tau \nonumber \\
&+\int_{\tau_1=0}^\infty\int_{\tau_2=0}^{y_1(\tau_1)-\tau_1} \bar{y}_2(\tau_1,\tau_2)e^{-\tau_1}e^{-\tau_2}d\tau_1d\tau_2
\end{align}

We use the above results to characterize the structure of the optimal policy for problem (\ref{opt_rnwl}) in the next subsection.

\subsection{Optimal Solution for Problem (\ref{opt_rnwl}): Threshold Policies}

We define the following parameterized problem to characterize the optimal solution of problem (\ref{opt_rnwl})
\begin{align} \label{opt_lmda}
p_2(\lambda)\triangleq\min_{x_1,y_1,y_2,\bar{y}_2}\quad&\mathbb{E}\left[R\right]-\lambda\mathbb{E}\left[L\right] \nonumber \\
\mbox{s.t.}~~~ \quad &x_1\geq0 \nonumber \\
&y_1(\tau)\geq\tau, \quad \forall \tau \nonumber \\
&y_2(\tau)\geq\tau, \quad \forall \tau \nonumber \\
&\bar{y}_2(\tau_1,\tau_2)\geq\tau_1+\tau_2, \quad \forall \tau_1,\tau_2
\end{align}
where the subscript $2$ in $p_2(\lambda)$ denotes the $B=2$ case that we consider here. This approach has also been used in \cite{sun-weiner}. We now have the following lemma.

\begin{lemma} \label{thm_lmda}
$p_2(\lambda)$ is decreasing in $\lambda$, and the optimal solution of problem (\ref{opt_rnwl}) is given by $\lambda^*$ that solves $p_2(\lambda^*)=0$.
\end{lemma}

\begin{Proof}
Let $\lambda_1>0$, and let the solution of problem (\ref{opt_lmda}) be given by $\left(x_1^{(1)},y_1^{(1)},y_2^{(1)},\bar{y}_2^{(1)}\right)$ for $\lambda=\lambda_1$, with the corresponding average area under the age curve in the epoch and the average epoch length given by $\mathbb{E}\left[R^{(1)}\right]$ and $\mathbb{E}\left[L^{(1)}\right]$, respectively. Now for some $\lambda_2>\lambda_1$, one can write
\begin{align}
p_2(\lambda_1)&=\mathbb{E}\left[R^{(1)}\right]-\lambda_1\mathbb{E}\left[L^{(1)}\right] \nonumber \\
&>\mathbb{E}\left[R^{(1)}\right]-\lambda_2\mathbb{E}\left[L^{(1)}\right] \nonumber \nonumber \\
&\geq p_2(\lambda_2).
\end{align}
where the last inequality follows since $\left(x_1^{(1)},y_1^{(1)},y_2^{(1)},\bar{y}_2^{(1)}\right)$ is also feasible in problem (\ref{opt_lmda}) for $\lambda=\lambda_2$.

Next, note that both problems (\ref{opt_lmda}) and (\ref{opt_rnwl}) have the same feasible set. In addition, if $p_2(\lambda)=0$, then the objective function of (\ref{opt_rnwl}) satisfies $\mathbb{E}\left[R\right]/\mathbb{E}\left[L\right]=\lambda$. Hence, the objective function of (\ref{opt_rnwl}) is minimized by minimizing $\lambda\geq0$ such that $p_2(\lambda)=0$. Finally, by the first part of lemma, there can only be one such $\lambda$, which we denote $\lambda^*$.
\end{Proof}

By Lemma~\ref{thm_lmda}, one can simply use a bisection method to find $\lambda^*$ that solves $p_2(\lambda^*)=0$. This $\lambda^*$ certainly exists since $p_2(0)>0$ and $\lim_{\lambda\rightarrow\infty}p_2(\lambda)=-\infty$. We focus on problem (\ref{opt_lmda}) in the rest of this subsection, for which we introduce the following Lagrangian \cite{boyd}
\begin{align}
\mathcal{L}=&\mathbb{E}\left[R\right]-\lambda\mathbb{E}\left[L\right]-\eta_1x_1-\int_0^\infty\gamma_1(\tau)\left(y_1(\tau)-\tau\right)d\tau \nonumber \\
&-\int_0^\infty\gamma_2(\tau)\left(y_2(\tau)-\tau\right)d\tau \nonumber \\
&- \int_0^\infty\int_0^\infty\bar{\gamma}_2(\tau_1,\tau_2)\left(\bar{y}_2(\tau_1,\tau_2)-\tau_1-\tau_2\right)d\tau_1d\tau_2
\end{align}
where $\eta_1$, $\gamma_1(\cdot)$, $\gamma_2(\cdot)$, $\bar{\gamma}_2(\cdot,\cdot)$ are Lagrange multipliers. Using (\ref{eq_exp_r}) and (\ref{eq_exp_l}), we take the (functional) derivative of the Lagrangian with respect to $\bar{y}_2(t_1,t_2)$ and equate it to $0$ to get
\begin{align}
\bar{y}_2(t_1,t_2)=\lambda+\frac{\bar{\gamma}_2(t_1,t_2)}{e^{-(t_1+t_2)}}
\end{align}
Now if $t_1+t_2<\lambda$, then $\bar{y}_2(t_1,t_2)$ has to be larger than $t_1+t_2$, for if it were equal, the right hand side of the above equation would be larger than the left hand side. By complementary slackness \cite{boyd}, we conclude that in this case $\bar{\gamma}_2(t_1,t_2)=0$, and hence $\bar{y}_2(t_1,t_2)=\lambda$. On the other hand, if $t_1+t_2\geq\lambda$, then $\bar{y}_2(t_1,t_2)$ has to be equal to $t_1+t_2$, for if it were larger, then by complementary slackness $\bar{\gamma}_2(t_1,t_2)=0$ and the right hand side of the above equation would be smaller than the left hand side. In conclusion, we have
\begin{align} \label{eq_y2_bar}
\bar{y}_2(t_1,t_2)=\begin{cases}\lambda,\quad &t_1+t_2<\lambda \\ t_1+t_2,\quad &t_1+t_2\geq\lambda\end{cases}
\end{align}
The above result says that starting from state $(0,0)$ the sensor has to wait at least for $\lambda$ time units before submitting an update, provided that it received two consecutive energy units (without using the first one to send an update) in that epoch. If these two energy arrivals occur relatively early, i.e., $t_1+t_2<\lambda$, then the sensor updates exactly after $\lambda$ time units from the beginning of the epoch. Otherwise, if $t_1+t_2\geq\lambda$, then the sensor updates instantly after receiving the second energy unit. We coin this type of policies {\it $\lambda$-threshold policy,} where the sensor can only update if the age grows above a certain threshold $\lambda$. Such policies were first introduced in the solution of the case of $B=1$ energy unit in \cite{jing-age-online}, and have also appeared in the random full battery recharges analysis in \cite{arafa-age-rbr}. We also note from the result in (\ref{eq_y2_bar}) that $\bar{y}_2(t_1,t_2)$ only depends on the age at the second energy arrival, $t_1+t_2$.

Next, we take the derivative of the Lagrangian with respect to $y_2(t)$ and equate to $0$ to get
\begin{align}
y_2(t)=\lambda+\frac{\gamma_2(t)}{qe^{x_1}e^{-t}}
\end{align}
where $q\triangleq1-\int_0^\infty e^{-y_1(\tau)}d\tau$. Note that $q\in[0,1]$ since $y_1(\tau)\geq\tau$. Following the same arguments as in the $\bar{y}_2$ case, we get that
\begin{align} \label{eq_y2}
y_2(t)=\begin{cases}\lambda,\quad &t<\lambda \\ t,\quad &t\geq\lambda\end{cases}
\end{align}
That is, $y_2$ is also a $\lambda$-threshold policy, and $\bar{y}_2(t_1,t_2)=y_2(t_1+t_2)$. This settles the earlier question we posed at the beginning of this section of whether receiving two energy arrivals starting from state $(0,0)$ would lead to a different policy than receiving one energy arrival starting from state $(1,0)$; the optimal policy when the sensor has a full battery is only a function of the age at the time of receiving the second energy unit in the battery. Next, we take the derivative of the Lagrangian with respect to $x_1$ and equate to $0$ to get
\begin{align} \label{eq_x1_ini}
x_1=&\lambda+e^{x_1}\int_0^{x_1}\left(\lambda y_2(\tau)-\frac{1}{2}y_2(\tau)^2\right)e^{-\tau}d\tau \nonumber \\
&\hspace{.75in}+\lambda y_2\left(x_1^-\right) - \frac{1}{2}y_2\left(x_1^-\right)^2 + \frac{\eta_1}{q}
\end{align}
We now make an assumption that $x_1>\lambda$, and verify that assumption below. Based on that, $y_2\left(x_1^-\right)=x_1$ from (\ref{eq_y2}). One can also use (\ref{eq_y2}) to evaluate the integral in the above equation in terms of $\lambda$ and $x_1$. After some algebraic manipulations, we get that for $x_1>0$, $\eta_1=0$ by complementary slackness, and the following holds
\begin{align} \label{eq_x1}
x_1=\log\left(\frac{1}{e^{-\lambda}-\frac{1}{2}\lambda^2}\right)
\end{align}
where $\log$ is the natural logarithm. It is direct to see from (\ref{eq_x1}) that $x_1>\lambda$ as assumed above.

Finally, we take the derivative of the Lagrangian with respect to $y_1(t)$ and equate to $0$ to get
\begin{align} \label{eq_y1_ini}
y_1(t)=&\lambda +e^{x_1}\!\!\int_0^{x_1}\!\!\!\left(\lambda y_2(\tau)-\frac{1}{2}y_2(\tau)^2\right)\!e^{-\tau}d\tau +\lambda x_1-\frac{1}{2}x_1^2 \nonumber \\
&+\frac{1}{2}y_1(t)^2 -\frac{1}{2}\bar{y}_2\left(t,\left(y_1(t)-t\right)^-\right)^2 \nonumber \\
&-\lambda y_1(t) +\lambda \bar{y}_2\left(t,\left(y_1(t)-t\right)^-\right) +\frac{\gamma_1(t)}{e^{-y_1(t)}}
\end{align}
We now make another assumption that $y_1(t)>\lambda,~\forall t$, and verify it below. Based on this assumption, we conclude by (\ref{eq_y2_bar}) that $\bar{y}_2\left(t,\left(y_1(t)-t\right)^-\right)=y_1(t)$. We substitute this in (\ref{eq_y1_ini}), and use (\ref{eq_x1_ini}) to get
\begin{align} \label{eq_y1_kkt}
y_1(t)=x_1+\frac{\gamma_1(t)}{e^{-y_1(t)}}
\end{align}
which verifies that $y_1(t)>\lambda,~\forall t$, since $x_1>\lambda$. Similar to the arguments used in deriving (\ref{eq_y2_bar}) and (\ref{eq_y2}), we conclude from (\ref{eq_y1_kkt}) that $y_1$ is an $x_1$-threshold policy given by
\begin{align} \label{eq_y1}
y_1(t)=\begin{cases}x_1,\quad&t<x_1 \\
t,\quad&t\geq x_1\end{cases}
\end{align}
Similar to the discussion regarding the equivalence of $\bar{y}_2$ and $y_2$, we conclude from (\ref{eq_y1}) that starting from state $(0,0)$ and receiving one energy unit is equivalent to starting from state $(1,0)$ and receiving no energy units; in both cases, the sensor has the same threshold $x_1$ after which it can update. Using (\ref{eq_y2_bar}), (\ref{eq_y2}), (\ref{eq_x1}), and (\ref{eq_y1}) we get that
\begin{align}
p_2(\lambda)=&\frac{1}{2}\lambda^2+(\lambda+1)e^{-\lambda}+\lambda \nonumber \\
&-\left(e^{-\lambda}-\frac{1}{2}\lambda^2+1\right)\log\left(\frac{1}{e^{-\lambda}-\frac{1}{2}\lambda^2}\right)
\end{align}

It now remains to find $\lambda^*$. Towards that, we first note that we have an upper bound on $\lambda^*$ given by $0.9012$, the solution of the $B=1$ case derived in \cite{jing-age-online}. We also have a lower bound of $0.5$, which is the optimal solution in the case of having an infinite battery, also derived in \cite{jing-age-online}. Using bisection, we find that the optimal solution at which $p_2(\lambda^*)=0$ is given by $\lambda^*\approx0.72$, with the corresponding $x_1^*\approx1.48$. Observe that the fact that $x_1^*$ is larger than $\lambda^*$ implies the intuitive behavior that the sensor is less eager to send an update if it has only one energy unit, compared to when it has a full battery of two energy units.

\subsection{Comparison to Other Policies}

\begin{figure}[t]
\center
\includegraphics[scale=.45]{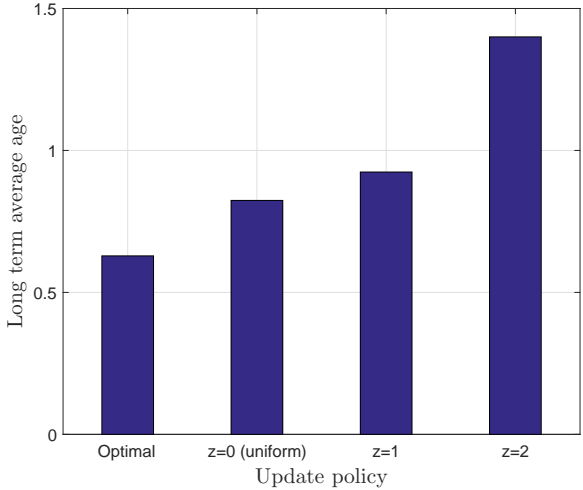}
\caption{Comparison of the optimal policy for $B=2$ to other policies: uniform updating, and energy-aware adaptive updating of \cite{jing-age-online}.}
\label{fig_age_cmpr}
\end{figure}

We now compare the optimal result derived above with other schemes and system models in the literature. We first compare it to the {\it energy-aware adaptive status update policy} introduced and analyzed in \cite{jing-age-online}. In there, the sensor schedules its next update based on the amount of energy in its battery; if the energy is less than $B/2$, it schedules the next update after $1/(1-\beta)$ time units, for some constant $\beta<1$; if the energy is larger than $B/2$, it schedules the next update after $1/(1+\beta)$ time units; and if the energy is exactly equal to $B/2$, it schedules the next update after $1$ time unit. Then, if the sensor has no energy at its scheduled update time, it stays silent, and reschedules its following update accordingly after $1/(1-\beta)$ time units. We note that for $\beta=0$, this energy-aware status update policy transforms into a {\it best effort uniform update policy,} which is the optimal solution for the infinite battery case \cite{jing-age-online}. We also note that the constant $\beta$ is chosen in \cite{jing-age-online} such that the policy is asymptotically optimal in the battery size. Specifically, it is chosen equal to $z\log{B}/B$ for some positive integer $z$ that controls the policy's asymptotic behavior. We compare our optimal policy to the energy-aware policy above for $z=2$, $z=1$, and $z=0$ (uniform update policy) in Fig.~\ref{fig_age_cmpr}. We see that it outperforms all of them.

Finally, we compare the optimal policy to our recent results on an altered system model of the same problem \cite{arafa-age-rbr}. There, the battery is {\it fully} recharged randomly over time, i.e., energy arrives in chunks of $B$ energy units, as opposed to the incremental unit recharges considered in this work. We consider two situations of this random battery recharges to compare with. The first is when the Poisson arrival process is of unit rate, and the second is when it is of rate $1/2$. The second case corresponds to an average recharge rate of $B/2=1$ energy unit per unit time, as considered in this work. From \cite{arafa-age-rbr}, the optimal long term average age for the first situation is given by $r_1^*=0.59$. While the analysis in \cite{arafa-age-rbr} is done for a Poisson arrival process of unit rate, it can be directly extended to account for that of rate $1/2$; this gives the optimal long term average age for the second situation by $r_2^*=1.18$, which is double $r_1^*$, since the average recharge rate is reduced to half. We conclude from this that while it is clearly better to have the battery recharged by $2$ energy units, as opposed to only $1$, every one time unit on average ($r_1^*<\lambda^*$), it is worse to be recharged by $2$ energy units every $2$ time units on average, as opposed to $1$ energy unit per unit time ($r_2^*>\lambda^*$), although the recharge rate is the same. The latter conclusion for the second situation is due to the fact that the system with $1$ energy unit recharge per unit time considered in this work gives more flexibility to the sensor on when to update compared to the system with $2$ energy units recharge every $2$ time units. This flexibility allows the sensor to submit updates more uniformly over time, which achieves better age by convexity of the square function that governs the areas of the triangles constituting the total area under the age curve to be minimized.

\section{Conclusion and Future Directions}

We have characterized optimal online policies for energy harvesting sensors with $B$-sized batteries that minimize the long term average age of information, subject to energy causality constraints. We have considered a noiseless channel where a transmission update consumes one energy unit and arrives instantaneously at the receiver. Under a Poisson energy arrival process with unit rate, energy units arrive at the sensor's battery in an incremental fashion, i.e., one energy unit per arrival. We first have shown that the optimal status update policy has a renewal structure. Specifically, the times between the two consecutive events of submitting an update and having $k$ energy units remaining in the battery afterwards, $0\leq k\leq B-1$, are i.i.d. Then, we have thoroughly studied the specific scenario of $B=2$ energy units and further shown that the optimal renewal policy has an {\it energy-dependent threshold} structure: the sensor submits an update only if the age of information surpasses a certain threshold which is a function of the energy available in its battery.

From the analysis of the $B=2$ case, it is amenable to show that threshold policies are also optimal for any $B\geq3$. One main difficulty in showing that is the combinatorial nature of how the different $B$ random variables that govern the energy arrivals in between inter-updates are related. Similar to the approaches in \cite{ozgur_online_su, ozgur_online_mac, baknina_online_mac, baknina_online_bc, baknina-online-proc, arafa-baknina-gnrl-online}, it is therefore of interest to study near-optimal renewal-type policies that provably perform within a constant gap from the optimal solution of problem (\ref{opt_main}) in future works.

\appendix

\subsection{Proof of Theorem~\ref{thm_rnwl}} \label{apndx_rnwl}

We prove this by showing that any given status update policy that is uniformly bounded according to Definition~\ref{def_ubp} is outperformed by a renewal policy as defined in the theorem. Let us consider the $i$th epoch (time between two consecutive visits to state $(k,0)$); we introduce the following notation regarding the energy arrivals occurring in it. Let $\tau_{1,i}$ denote the time until the first energy arrival after the epoch starts, and let there be $j_1$ status updates after that energy arrival {\it before} a second energy arrival occurs. If $j_1\geq1$, then let $\tau_{2,i}$ denote the time until the first energy arrival {\it after} the $j_1$th update. Otherwise, if $j_1=0$, then let $\tau_{2,i}$ denote the inter-arrival time between the first and the second energy arrivals in the epoch. Similarly, let there be $j_2$ status updates after the second energy arrival {\it before} a third energy arrival occurs. If $j_2\geq1$, then let $\tau_{3,i}$ denote the time until the first energy arrival {\it after} the $j_2$th update. Otherwise, if $j_2=0$, then let $\tau_{3,i}$ denote the inter-arrival time between the second and the third energy arrivals in the epoch. We continue defining $\tau_{j,i}$'s, $j=1,2,\dots$, until the epoch ends by retuning back to state $(k,0)$ again. Finally, in the event that the $j$th energy arrival in the epoch makes the battery full, then we wait until the first status update occurs after that event and denote by $\tau_{j+1,i}$ the time until the first energy arrival {\it after} that update, i.e., we do not account for energy arrivals that cause battery overflows.

As noted before Theorem~\ref{thm_rnwl}, there can possibly be an infinite number of updates before the system returns back to state $(k,0)$, depending on the energy arrival pattern and the update time decisions. For a given status update policy, one can enumerate all such patterns. For instance, following the above notation, the first pattern could be when the system goes from state $(k,0)$ to state $(k+1,\tau_{1,i})$ and then to state $(k,0)$ again; the second pattern could be when the system goes through the following sequence of states: $(k,0)-(k+1,\tau_{1,i})-(k+2,\tau_{1,i}+\tau_{2,i})-(k+1,0)-(k,0)$; and so on. Let the vector ${\bm \tau}_{m,i}$ contain all the $\tau_{j,i}$'s in the $m$th pattern. Note that this vector's length varies with the pattern. For instance, we have ${\bm \tau}_{1,i}=\tau_{1,i}$ and ${\bm \tau}_{2,i}=[\tau_{1,i},\tau_{2,i}]$ for the above two pattern examples, respectively. For a given status update policy, one can also compute the probability of occurrence of the $m$th pattern in the $i$th epoch, denoted by $p_{m,i}$, with $\sum_{m=1}^\infty p_{m,i}=1$. Let us also denote by $R_{m,i}$ the area under the age curve in that epoch, given that it went through the $m$th pattern.

Next, for a fixed history $\mathcal{H}_{i-1}$ and a pattern $m$, let us group all the status updating sample paths that have the same ${\bm \tau}_{m,i}$ and perform a statistical averaging over all of them to get the following average age in the $i$th epoch given that it went through the $m$th pattern
\begin{align}
\hat{R}_{m,i}\left({\bm \gamma}_m,\mathcal{H}_{i-1}\right)\triangleq\mathbb{E}\left[R_{m,i}|{\bm \tau}_{m,i}={\bm \gamma}_m,\mathcal{H}_{i-1}\right]
\end{align}
Now for a given time $T$, let $N_T$ denote the number of epochs that have already {\it started} by time $T$. Then, we have
\begin{align}
\mathbb{E}&\left[R_{m,i}\cdot\mathbbm{1}_{i\leq N_T}\right] \nonumber \\
&=\mathbb{E}_{\mathcal{H}_{i-1}}\left[\mathbb{E}_{{\bm \tau}_{m,i}}\left[\hat{R}_{m,i}\left({\bm \gamma}_m,\mathcal{H}_{i-1}\right)\right]\cdot\mathbbm{1}_{i\leq N_T}\Big|\mathcal{H}_{i-1}\right] \label{eq_pf_ren_1}
\end{align}
where equality follows since $\mathbbm{1}_{i\leq N_T}$ is independent of ${\bm \tau}_{m,i}$ given $\mathcal{H}_{i-1}$. Similarly, let $x_{k,m,i}$ denote the length of the $i$th epoch under the $m$th pattern, and define its (conditional) average as
\begin{align}
\hat{x}_{k,m,i}\left({\bm \gamma}_m,\mathcal{H}_{i-1}\right)\triangleq\mathbb{E}\left[x_{k,m,i}|{\bm \tau}_{m,i}={\bm \gamma}_m,\mathcal{H}_{i-1}\right]
\end{align}
Finally, we denote by $R_i$ and $x_{k,i}$ the area under the age curve in the $i$th epoch and its length, respectively, irrespective of which pattern it went through.

\begin{figure*}[t]
\begin{align}
\frac{\mathbb{E}\left[\sum_{i=1}^\infty R_i\mathbbm{1}_{i\leq N_T}\right]}{\mathbb{E}\left[\sum_{i=1}^\infty x_{k,i}\mathbbm{1}_{i\leq N_T}\right]} &=\frac{\sum_{i=1}^\infty \mathbb{E}_{\mathcal{H}_{i-1}}\left[ \sum_{m=1}^\infty p_{m,i} \mathbb{E}_{{\bm \tau}_{m,i}}\left[\hat{R}_{m,i}\left({\bm \gamma}_m,\mathcal{H}_{i-1}\right)\right]\cdot\mathbbm{1}_{i\leq N_T}\Big|\mathcal{H}_{i-1}\right]}{\mathbb{E}\left[\sum_{i=1}^\infty x_{k,i}\mathbbm{1}_{i\leq N_T}\right]} \label{eq_pf_ren_2} \\
&=\frac{\sum_{i=1}^\infty \mathbb{E}_{\mathcal{H}_{i-1}}\left[ \sum_{m=1}^\infty p_{m,i} \mathbb{E}_{{\bm \tau}_{m,i}}\left[\hat{x}_{k,m,i}\left({\bm \gamma}_m,\mathcal{H}_{i-1}\right)\right]
\cdot \frac{\sum_{m=1}^\infty p_{m,i}\mathbb{E}_{{\bm \tau}_{m,i}}\left[\hat{R}_{m,i}\left({\bm \gamma}_m,\mathcal{H}_{i-1}\right)\right]}{\sum_{m=1}^\infty p_{m,i}\mathbb{E}_{{\bm \tau}_{m,i}}\left[\hat{x}_{k,m,i}\left({\bm \gamma}_m,\mathcal{H}_{i-1}\right)\right]} 
\cdot \mathbbm{1}_{i\leq N_T}\Big|\mathcal{H}_{i-1}\right]}{\mathbb{E}\left[\sum_{i=1}^\infty x_{k,i}\mathbbm{1}_{i\leq N_T}\right]} \\
&\geq\frac{\sum_{i=1}^\infty \mathbb{E}_{\mathcal{H}_{i-1}}\left[ \sum_{m=1}^\infty p_{m,i} \mathbb{E}_{\tau_i}\left[\hat{x}_{k,m,i}\left(\gamma,\mathcal{H}_{i-1}\right)\right]
\cdot R^*\left(\mathcal{H}_{i-1}\right)
\cdot \mathbbm{1}_{i\leq N_T}\Big|\mathcal{H}_{i-1}\right]}{\mathbb{E}\left[\sum_{i=1}^\infty x_{k,i}\mathbbm{1}_{i\leq N_T}\right]} \\
&\geq R_{\min}
\end{align}
\hrulefill
\end{figure*}

Next, note that by (\ref{eq_aoi}), the following holds
\begin{align} \label{eq_aoi_bd}
\frac{1}{T}\sum_{i=1}^\infty R_i\mathbbm{1}_{i\leq N_T-1}\leq \frac{r(T)}{T} \leq \frac{1}{T}\sum_{i=1}^\infty R_i\mathbbm{1}_{i\leq N_T}
\end{align} 
Following similar analysis as in \cite[Appendix C-1]{jing-age-online}, one can show that
\begin{align}
\lim_{T\rightarrow\infty} \frac{\mathbb{E}\left[R_{N_T}\right]}{T}=0
\end{align}
for any uniformly bounded policy as in Definition~\ref{def_ubp}. Hence, the expected values of the upper and lower bounds in (\ref{eq_aoi_bd}) are equal as $T\rightarrow\infty$. Hence, in the sequel, we derive a lower bound on $\frac{1}{T}\mathbb{E}\left[\sum_{i=1}^\infty R_i\mathbbm{1}_{i\leq N_T}\right]$ and use the above note to conclude that it is also a lower bound on $\frac{\mathbb{E}\left[r(T)\right]}{T}$ as $T\rightarrow\infty$. Towards that end, note that $\mathbb{E}\left[\sum_{i=1}^\infty x_{k,i}\mathbbm{1}_{i\leq N_T}\right]\geq T$. Then, we have
\begin{align}
\frac{1}{T}\mathbb{E}\left[\sum_{i=1}^\infty R_i\mathbbm{1}_{i\leq N_T}\right] \geq \frac{\mathbb{E}\left[\sum_{i=1}^\infty R_i\mathbbm{1}_{i\leq N_T}\right]}{\mathbb{E}\left[\sum_{i=1}^\infty x_{k,i}\mathbbm{1}_{i\leq N_T}\right]}
\end{align}
We now proceed by lower bounding the right hand side of the above equation through a series of equations at the top of the next page. In there, (\ref{eq_pf_ren_2}) follows from (\ref{eq_pf_ren_1}) and the monotone convergence theorem, together with the fact that $\mathbb{E}\left[R_i\right]=\sum_{m=1}^\infty p_{m,i}\mathbb{E}\left[R_{m,i}\right]$; $R^*\left(\mathcal{H}_{i-1}\right)$ is the minimum value of $\frac{\sum_{m=1}^\infty p_{m,i}\mathbb{E}_{{\bm \tau}_{m,i}}\left[\hat{R}_{m,i}\left({\bm \gamma}_m,\mathcal{H}_{i-1}\right)\right]}{\sum_{m=1}^\infty p_{m,i}\mathbb{E}_{{\bm \tau}_{m,i}}\left[\hat{x}_{k,m,i}\left({\bm \gamma}_m,\mathcal{H}_{i-1}\right)\right]}$; and $R_{\min}$ is the minimum value of $R^*\left(\mathcal{H}_{i-1}\right)$ over all possible epochs and their corresponding histories, i.e., the minimum over all $i$ and $\mathcal{H}_{i-1}$. This, together with the fact that $\mathbb{E}\left[x_{k,i}\right]=\sum_{m=1}^\infty p_{m,i}\mathbb{E}\left[x_{k,m,i}\right]$, gives the last inequality.

Observe that a policy achieving $R^*\left(\mathcal{H}_{i-1}\right)$ is a policy which is a function of the possible energy arrival patterns in the $i$th epoch ${\bm \tau}_{m,i}$'s only, since the history $\mathcal{H}_{i-1}$ is fixed. Since the energy arrival process is Poisson with rate $1$, it follows that the random vector ${\bm \tau}_{m,i}$ consists of i.i.d. exponential random variables with parameter $1$, and that $\{{\bm \tau}_{m,i}\}$ are also independent across epochs. Therefore, if we repeat the policy that achieves $R_{\min}$ over all epochs, we get a renewal policy where the epoch lengths are also i.i.d., and $\{l_i\}$ forms a renewal process. This completes the proof.

\subsection{Justification of (\ref{eq_exp_r})} \label{apndx_jstfy}

First, we have
\begin{align} \label{eq_exp_r_1}
\mathbb{E}&\left[\frac{1}{2}y_1(\tau_1)^2\mathbbm{1}_{\tau_2>y_1(\tau_1)-\tau_1}\right] \nonumber\\
&=\int_{\tau_1=0}^\infty\int_{\tau_2=y_1(\tau_1)-\tau_1}^\infty\frac{1}{2}y_1(\tau_1)^2e^{-\tau_1}e^{-\tau_2}d\tau_2d\tau_1 \nonumber\\
&=\int_{\tau_1=0}^\infty\frac{1}{2}y_1(\tau_1)^2e^{-\tau_1}e^{-(y_1(\tau_1)-\tau_1)}d\tau_1 \nonumber\\
&=\int_{\tau_1=0}^\infty\frac{1}{2}y_1(\tau_1)^2e^{-y_1(\tau_1)}d\tau_1
\end{align}
Next, let us define
\begin{align}
\alpha\triangleq\mathbb{E}\left[\left(\frac{1}{2}\bar{y}_2(\tau_1,\tau_2)^2+\frac{1}{2}x_1^2\right)\mathbbm{1}_{\tau_2\leq y_1(\tau_1)-\tau_1}\right]
\end{align}
Since $\tau_m$ is independent of $\tau_1$ and $\tau_2$ for $m\geq3$, and they are all i.i.d., we have that the term $\alpha$ gets multiplied by $\sum_{i=1}^\infty(1-e^{-x_1})^{i-1}e^{-x_1}=1$ when we compute $\mathbb{E}[R]$. Note that
\begin{align}
\mathbb{E}\left[\mathbbm{1}_{\tau_2\leq y_1(\tau_1)-\tau_1}\right]&=1-\mathbb{P}\left[\tau_2> y_1(\tau_1)-\tau_1\right] \nonumber\\
&=1-\int_{\tau_1=0}^\infty\int_{\tau_2=y_1(\tau_1)-\tau_1}^\infty e^{-\tau_1}e^{-\tau_2}d\tau_2d\tau_1 \nonumber\\
&=1-\int_{\tau_1=0}^\infty e^{-\tau_1}e^{-(y_1(\tau_1)-\tau_1)}d\tau_1 \nonumber\\
&=1-\int_{\tau_1=0}^\infty e^{-y_1(\tau_1)}d\tau_1
\end{align}
and hence the term $\alpha$ can be expanded to
\begin{align} \label{eq_exp_r_2}
\alpha=&\frac{1}{2}x_1^2\left(1-\int_{\tau_1=0}^\infty e^{-y_1(\tau_1)}d\tau_1\right) \nonumber\\
&+\int_{\tau_1=0}^\infty\int_{\tau_2=0}^{y_1(\tau_1)-\tau_1}\frac{1}{2}\bar{y}_2(\tau_1,\tau_2)^2e^{-\tau_1}e^{-\tau_2}d\tau_1d\tau_2
\end{align}
Next, let us define
\begin{align}
\beta_m\triangleq\int_0^{x_1}\frac{1}{2}y_2(\tau_m)^2e^{-\tau_m}d\tau_m,\quad m\geq3
\end{align}
Now observe that, again by the fact that $\tau_i$'s are i.i.d., the terms $\beta_m$'s appear as follows when we take $\mathbb{E}[R]$

\begin{align} \label{eq_exp_r_3}
&\beta_3\mathbb{E}\left[\mathbbm{1}_{\tau_2\leq y_1(\tau_1)-\tau_1}\right]+\sum_{m=4}^\infty\!\beta_m\mathbb{E}\left[\mathbbm{1}_{\tau_2\leq y_1(\tau_1)-\tau_1}\right]\!\prod_{i=3}^m\mathbb{E}\left[\mathbbm{1}_{\tau_i\leq x_1}\right] \nonumber\\
&\quad=\beta_3\mathbb{E}\left[\mathbbm{1}_{\tau_2\leq y_1(\tau_1)-\tau_1}\right]\sum_{i=0}^\infty\left(1-e^{-x_1}\right)^i \nonumber\\
&\quad=e^{x_1}\beta_3\left(1-\int_{\tau_1=0}^\infty e^{-y_1(\tau_1)}d\tau_1\right)
\end{align}
where the second equality follows since $\beta_m$ is the same for all $m$. Equations (\ref{eq_exp_r_1}), (\ref{eq_exp_r_2}), and (\ref{eq_exp_r_3}) yield $\mathbb{E}[R]$ in (\ref{eq_exp_r}).


\end{document}